%% file: rig_space_neural_rendering.tex
\documentclass{article}

\usepackage{arxiv}

\usepackage[utf8]{inputenc} % allow utf-8 input
\usepackage[T1]{fontenc}    % use 8-bit T1 fonts
\usepackage{hyperref}       % hyperlinks
\usepackage{url}            % simple URL typesetting
\usepackage{booktabs}       % professional-quality tables
\usepackage{amsfonts}       % blackboard math symbols
\usepackage{nicefrac}       % compact symbols for 1/2, etc.
\usepackage{microtype}      % microtypography
\usepackage{lipsum}
\usepackage{graphicx}
\graphicspath{ {./images/} }

\usepackage{amssymb}
\usepackage{amsmath}
\usepackage{mathtools}

\usepackage{wrapfig}
\usepackage{multirow}
\usepackage{textcomp}
\usepackage{color}
\usepackage{float}

\usepackage{subfigure}

\newcommand{\comment}[1]{}

\title{Rig-space Neural Rendering}

\author{
Dominik Borer \\
  ETH Zurich \\
  Disney Research Studios \\
  \texttt{dborer@inf.ethz.ch} \\
   \And
Lu Yuhang \\
  ETH Zurich\\
  \texttt{luyuh@student.ethz.ch} \\
  \And
 Laura Wuelfroth \\
  ETH Zurich\\
  \texttt{wlaura@student.ethz.ch} \\
    \And
 Jakob Buhmann \\
  Disney Research Studios\\
  \texttt{jakob.buhmann@disneyresearch.com} \\
      \And
 Martin Guay \\
  Disney Research Studios\\
  \texttt{martin.guay@disneyresearch.com} \\
}

\begin{document}
%\maketitle

\input{fig/Figures.tex}

\maketitle

\TeaserFigNewColor

\begin{abstract}

\input{sec/Abstract.tex}

\end{abstract}

\input{sec/Introduction.tex}

\input{sec/RelatedWork.tex}
\input{sec/Dataset.tex}

\input{sec/Network.tex}

\input{sec/SceneComposition.tex}

\input{sec/Results.tex}

\input{sec/Limitations.tex}

\input{sec/Conclusion.tex}

\bibliographystyle{unsrt}  
\bibliography{rig_space_neural_rendering}  %%% Remove comment to use the external .bib file (using bibtex).
%%% and comment out the ``thebibliography'' section.

%%% Comment out this section when you \bibliography{references} is enabled.

\end{document}

%% file: fig/Figures.tex
\newcommand{\TeaserFigNewColor}{
\begin{figure*}[h]
\centering
\includegraphics[width=0.98\linewidth]{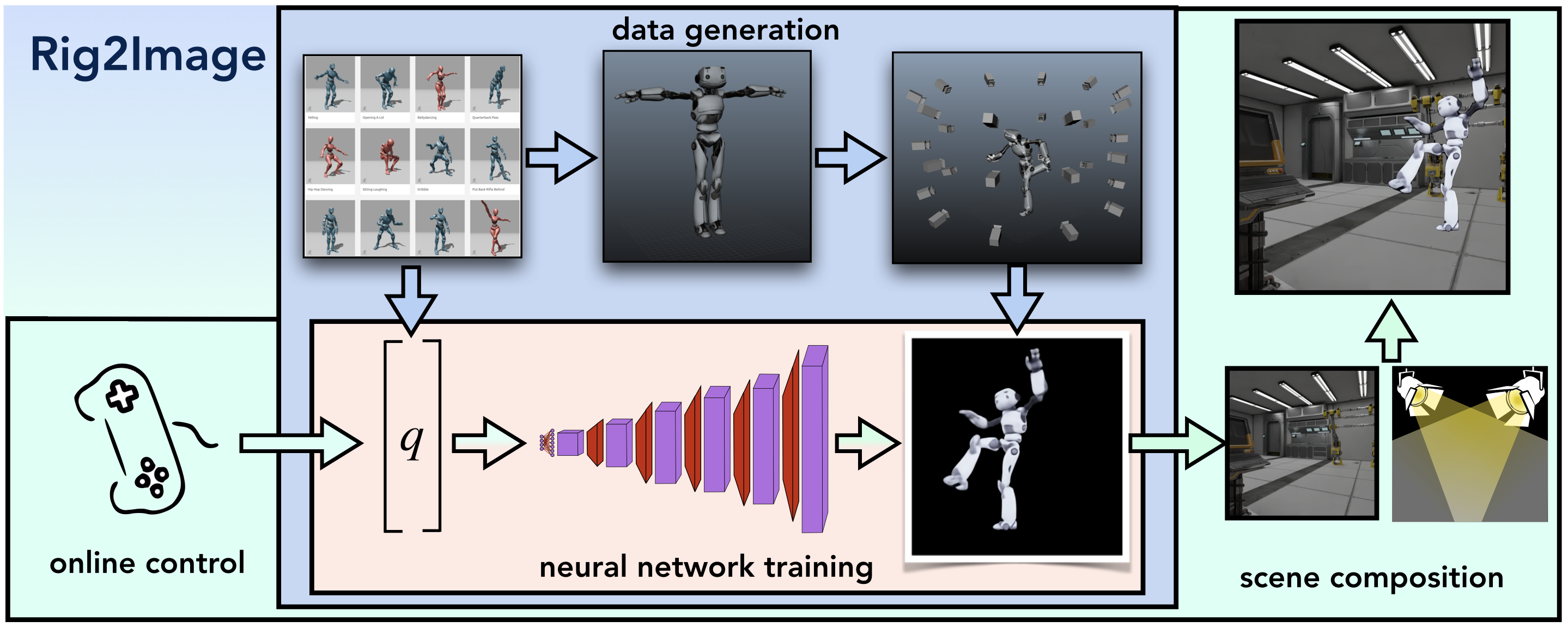}
\caption{With rig-space neural rendering, we train a deep neural network (DNN) to generate an image of a character directly from rig parameters, such as skeleton joint angles, or lattice cage positions. We first generate the training set (blue area) by posing the character in many poses and views, while saving the rig parameters associated to the images. We then train a DNN to generate the character image from the rig parameters (red area). Finally, at run-time (green area) we control the rig parameters and feed them to our \emph{Rig2Image} network that generates an image in real-time, which we then overlay onto scene background. Details on how to extend this to dynamic lights and interactions with other scene objects are shown in figures bellow.}
\label{Teaser}
\end{figure*}
}

\newcommand{\FigRuntime}{
	\begin{figure*}[h]
		\centering
		\includegraphics[width=0.98\linewidth]{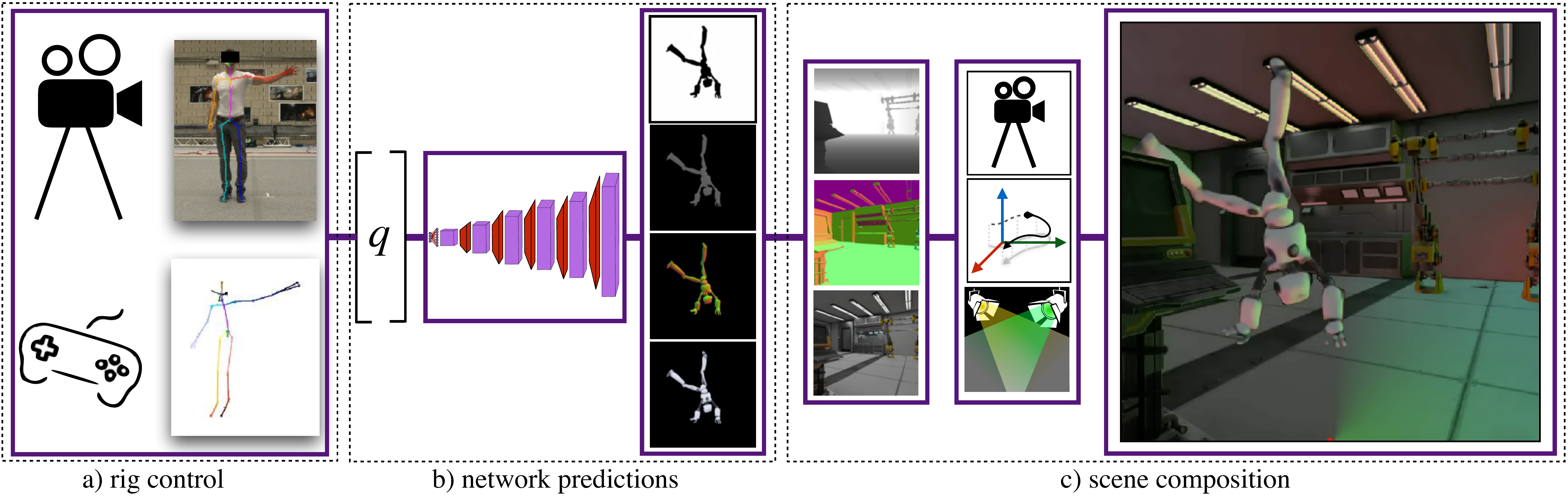}
		\caption{ \textbf{a)} Real-time applications control a skeleton or rig, as usual by blending skeletons, or by live capturing a human performer. \textbf{b)} The skeleton is then fed to our Rig2Image network that renders scene maps in real-time. \textbf{c) } These scene maps are then composed with dynamic lights as well as other scene objects in front and behind via the depth map, resulting in a final coherent scene.
		}
		\label{Runtime}
	\end{figure*}
}

\newcommand{\FigArchitecture}{
	\begin{figure*}[h]
		\centering
		\includegraphics[width=0.85\linewidth]{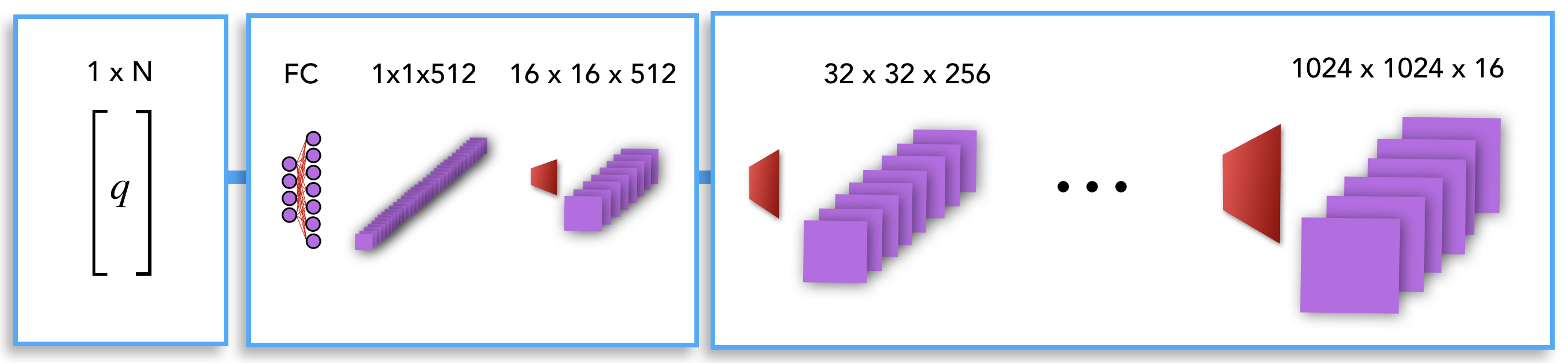}
		\caption{Our \emph{Rig2Image} architecture for high resolution images from rig parameters. The network takes as input rig parameters and first maps them to a fixed size ($512$) latent vector via a fully connected layer. These $1 \times 1 \times 512$ features then go through multiple convolutional blocks which double the dimensionality of the features, up to $1024\times1024$. To converge, we employ a training curriculum: we first train the \emph{backbone} of the network up to an image resolution of $16\times16$, then we train the convolutional blocks that increase the resolution, one at a time. }
		\label{Architecture}
	\end{figure*}
}

\newcommand{\FigExtensions}{
	\begin{figure*}[h]
	\centering
		\includegraphics[width=0.9\linewidth]{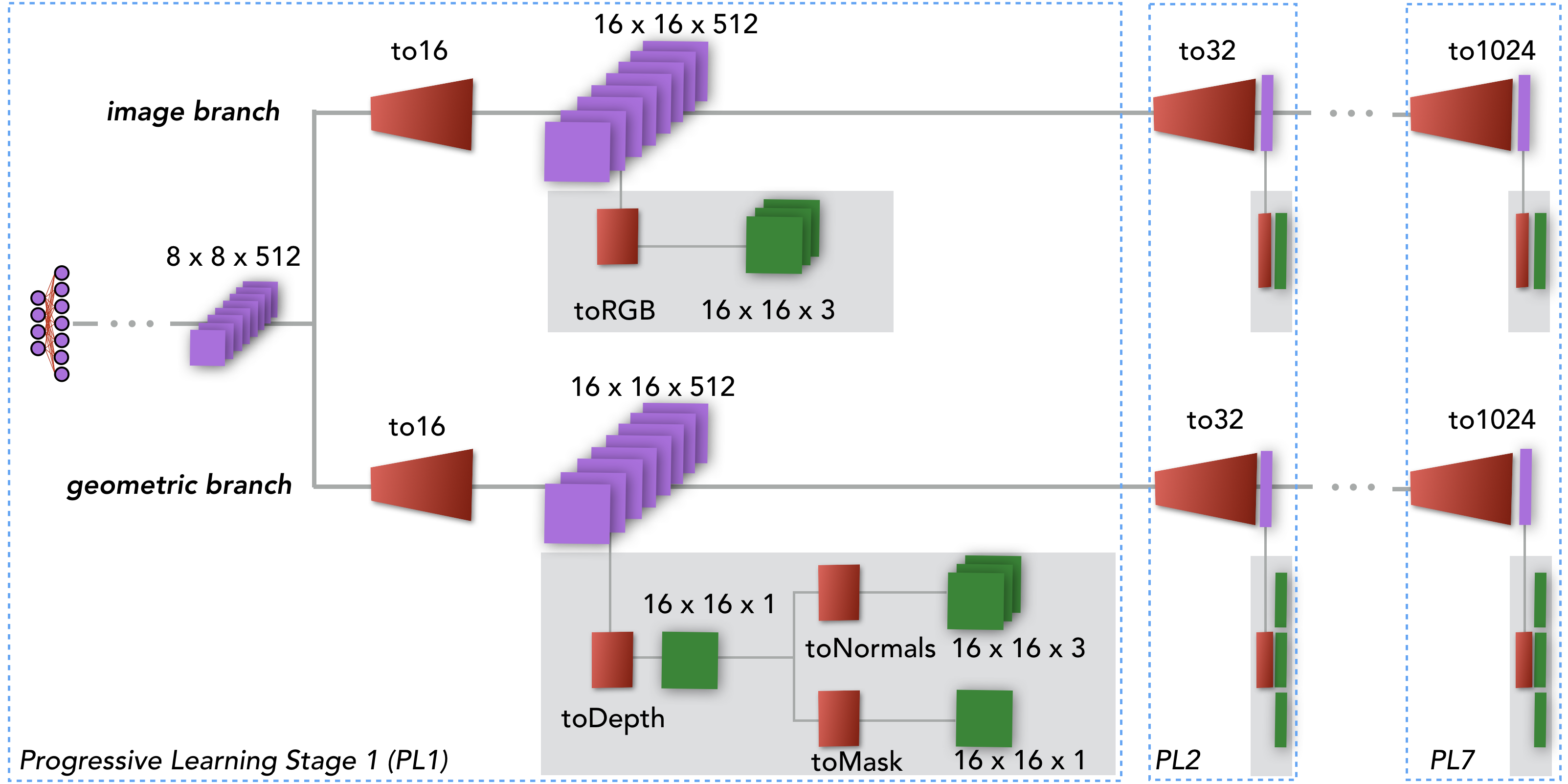}
		\caption{To predict feature maps such depth, normals, and mask, which are required for dynamic re-lighting and composition, we extend the network architecture to two main branches: an \emph{image} branch for albedo, and a \emph{geometric} branch for strongly correlated features such depth, normals and mask. Layers generating features up to $8\times8$ are shared, then split to produce higher resolutions. As before, we train the \emph{backbone} part of the network up to $16\times16$ resolution, referred to here as $PL1$. Subsequent blocks $PLi$ in each branch progress in parallel, trained one at a time, where the respective output layers (\emph{toRGB, toDepth, toNormals, toMask}) are only active at the currently trained resolution.}
		\label{Extensions}
	\end{figure*}
}

\newcommand{\FigResults}{
	\begin{figure*}[h]
	\centering
		\includegraphics[width=0.88\linewidth]{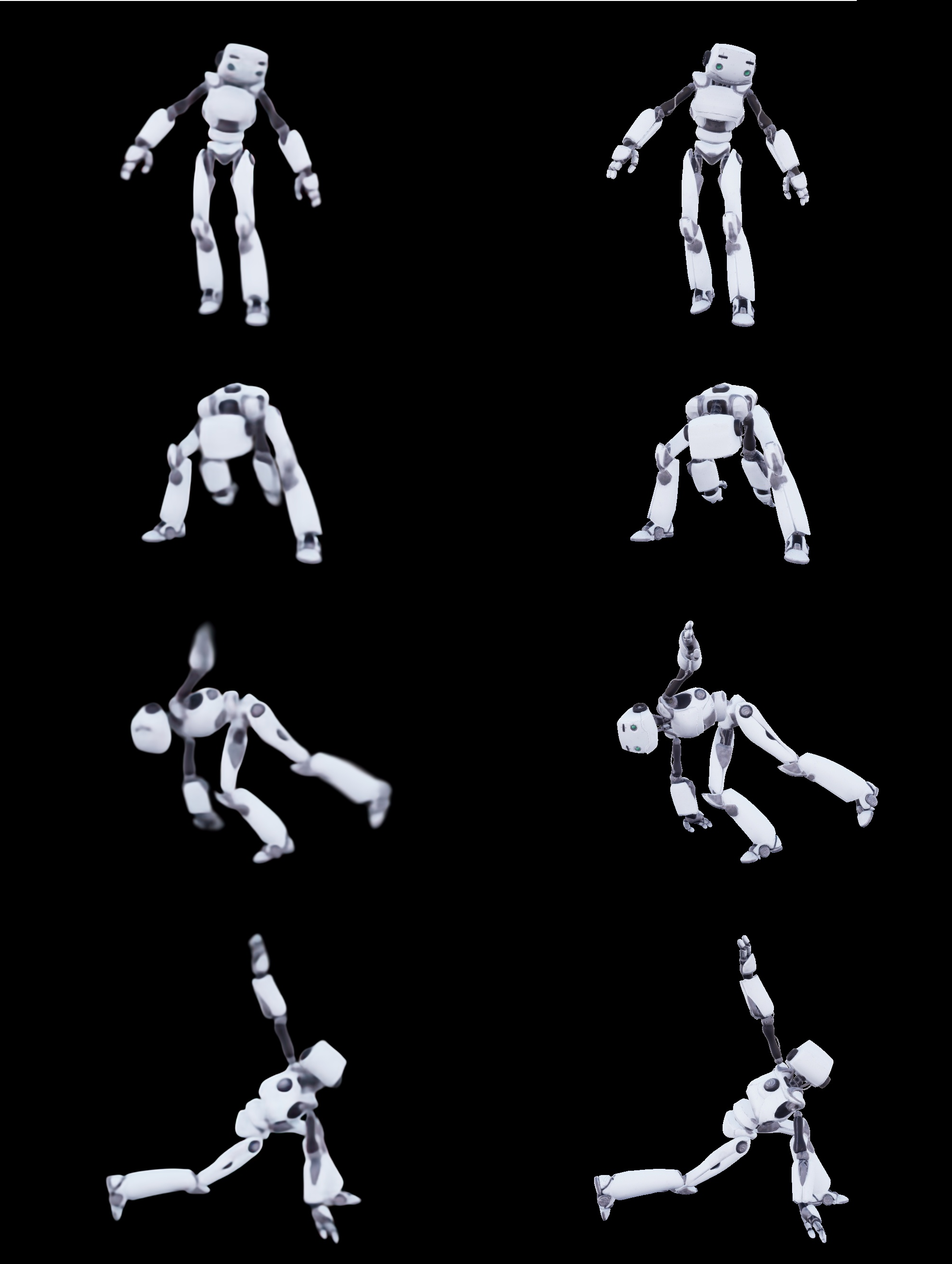}
		\caption{Results of our generated images from rig parameters. On the left is the generated image, while on the right is ground truth. We picked poses to illustrate the various motion clips we trained with. They include stretching movements, cartwheel rolls and yoga poses.  }
		\label{Results}
	\end{figure*}  
}

\newcommand{\FigFrontBack}{
	\begin{figure}[!h]
	\centering
		\includegraphics[width=0.95\columnwidth]{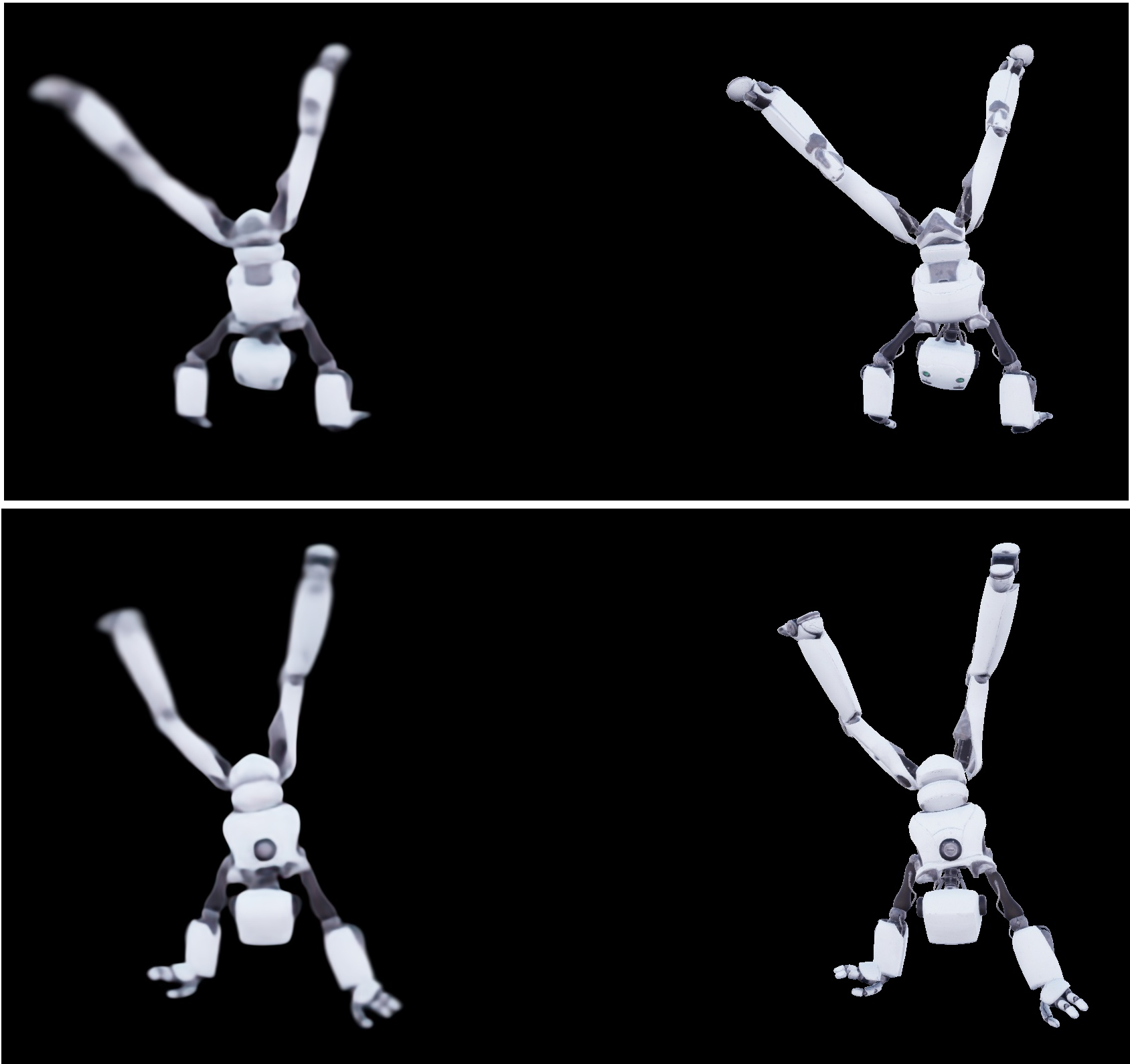}
		\caption{We can see that 3d rig parameters such as skeleton joint angles are sufficient to learn a disentangled surface map. The top row shows the character front facing, while the bottom row is the character back facing. }
		\label{frontback}
	\end{figure}
}

\newcommand{\FigNewViews}{
	\begin{figure}[!h]
	\centering
		\includegraphics[width=0.5\columnwidth]{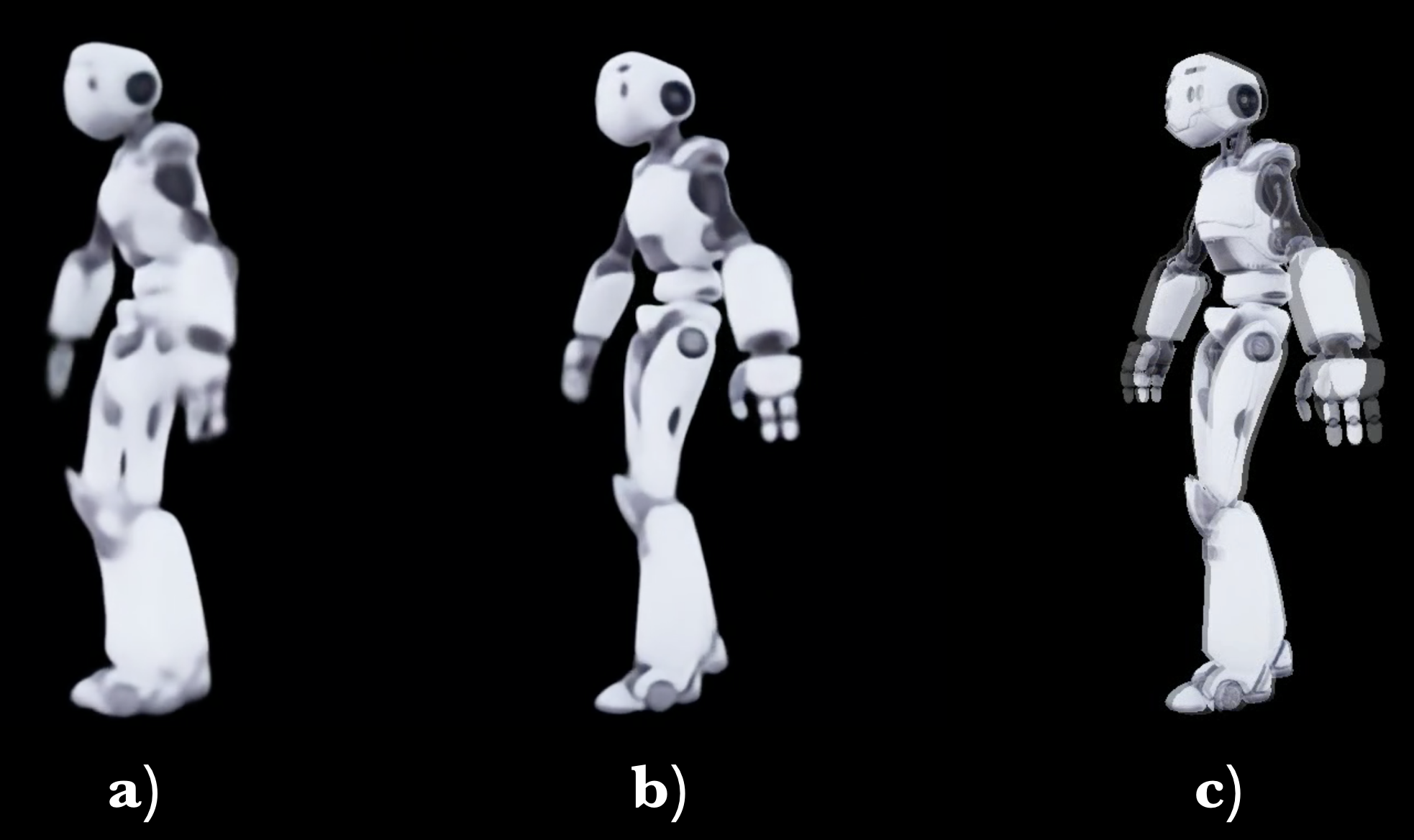}
		\caption{Comparison of results for novel view points: a) our model trained on $6$ views, b) our model trained on $60$ views, and c) a standard linear interpolation from the a ground truth dataset with $60$ viewpoint angles. }
		\label{NewViews}
	\end{figure}
}

\newcommand{\FigNewViewsDiagramm}{
	\begin{figure}[!h]
	\centering
		\includegraphics[width=0.5\columnwidth]{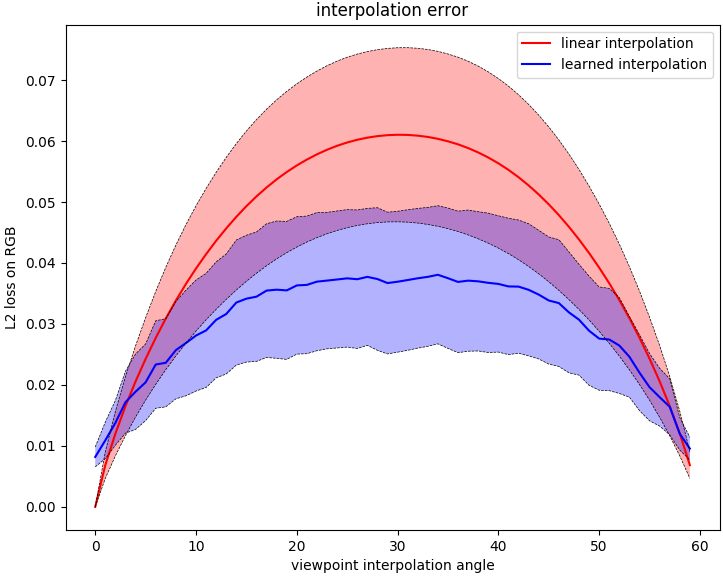}
		\caption{Quantitative error for the learned rendering over new views in blue, and linear interpolation in red. Both lines reflect the mean over 140 poses and the shaded region the standard deviation. The learned model was trained on $6$ views, hence the testing interval of $60^{\circ}$.}
		\label{fig_NewViewsDiagramm}
	\end{figure}
}

\newcommand{\FigResultsAllMaps}{

	\begin{figure}[t]
	\centering
		\includegraphics[width=0.7\linewidth]{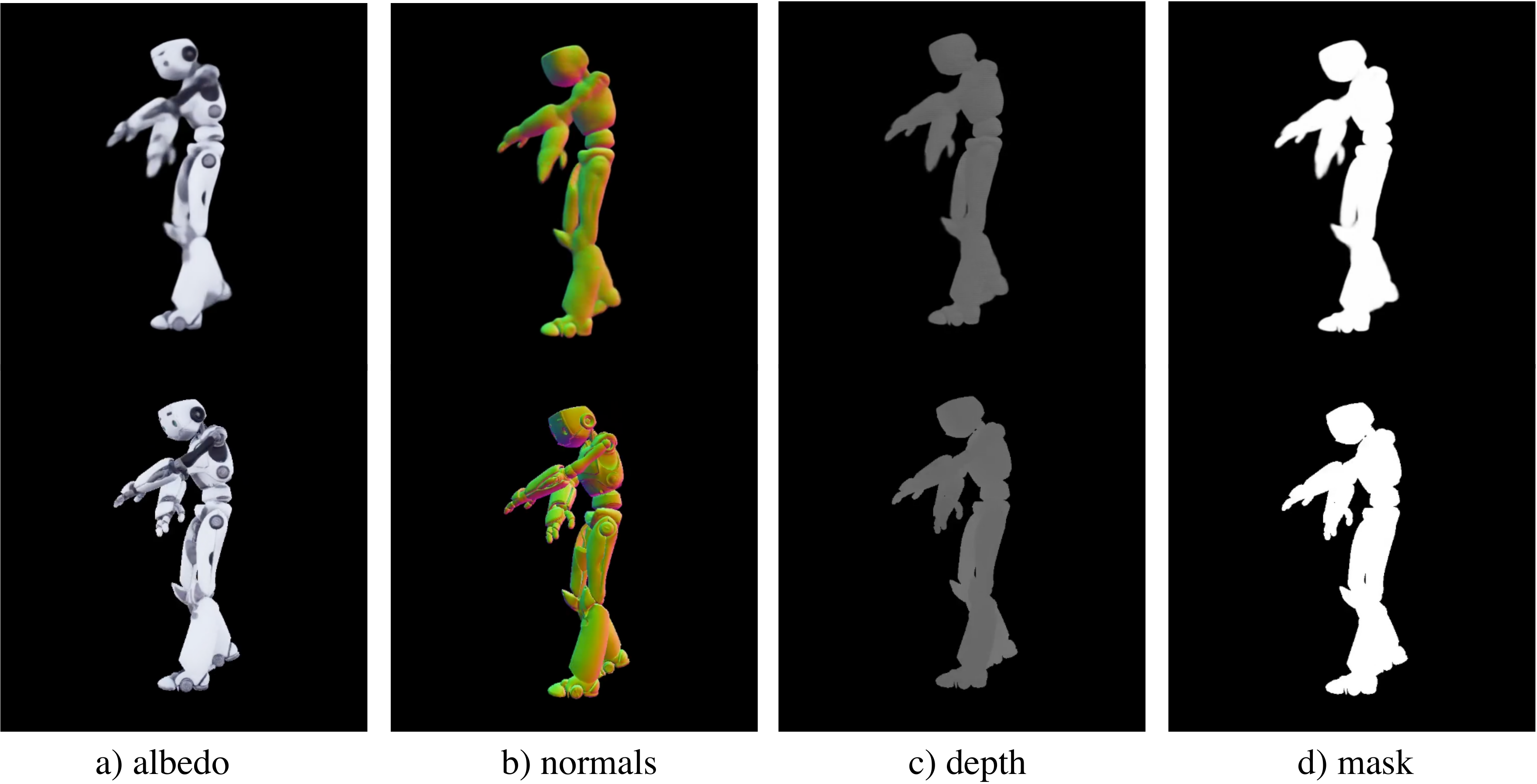}
		\caption{Example predictions (top) for the different scene maps and their according ground truth images (bottom). }
		\label{fig_results_all_maps}
	\end{figure}  
}

\newcommand{\FigResultWithError}{
	\begin{figure}[!h]
	\centering
		\includegraphics[width=0.95\linewidth]{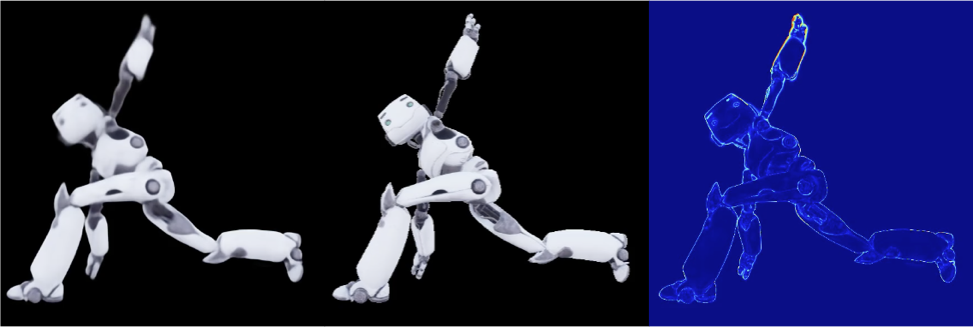}
		\caption{Example albedo output of the network: (left) prediction, (middle) ground truth, and (right) error visualization. As can be seen from the error plot, the system struggles mostly with high frequency details, such as the fine line of the mouth.}
		\label{fig_results_errror_vis}
	\end{figure}  
}

%% file: sec/Abstract.tex
Movie productions use high resolution 3d characters with complex proprietary rigs to create the highest quality images possible for large displays. Unfortunately, these 3d assets are typically not compatible with real-time graphics engines used for games, mixed reality and real-time pre-visualization. Consequently, the 3d characters need to be re-modeled and re-rigged for these new applications, requiring weeks of work and artistic approval. Our solution to this problem is to learn a compact image-based rendering of the original 3d character, conditioned directly on the rig parameters. Our idea is to render the character in many different poses and views, and to train a deep neural network to render high resolution images,  from the rig parameters directly. Many neural rendering techniques have been proposed to render from 2d skeletons, or geometry and UV. However these require manual work, and to do not remain compatible with the animator workflow of manipulating rig widgets, as well as the real-time game engine pipeline of interpolating rig parameters. We extend our architecture to support dynamic re-lighting and composition with other 3d objects in the scene. We designed a network that efficiently generates multiple scene feature maps such as normals, depth, albedo and mask, which are composed with other scene objects to form the final image.

%% file: sec/Introduction.tex
\section{Introduction}

Feature movie productions use high resolution characters with complex proprietary rigs to create the highest quality possible images for large displays. It is very often the case that these characters would be wanted and useful in a real-time setting, such as for previs, games or augmented reality experiences. Unfortunately, the high resolution assets cannot be rendered in real-time and the production rigs are often not easily transferable to game engine deformers (i.e. linear blend skinning and blend shapes). In consequence, characters must re-created manually including re-modeling their geometry with fewer polygons, re-creating textures with fewer details and re-rigging and re-animating their movements.  This process can be longer than expected, as the initial artistic intent of the character may be lost in one or several of these steps. 

%\FigTest
%\TeaserFigNewCOlor

%Since high resolution assets are not compatible with real-time engines, they must be re-created entirely---a process which requires weeks of manual work and artistic iteration. The characters are re-modelled with fewer polygons, re-textured with lower resolution materials, and re-rigged with skeletal linear blend skinning and blend shapes only. This reduction in resolution alters the initial artistic intent behind the characters.

\FigRuntime

\textbf{Our solution} to this problem is to treat the rendering of the character as a learned image-based rendering task, conditioned on 3d rig parameters such as a kinematic skeleton, blend shapes, or coarse mesh vertices. Our approach consists in rendering the 3d character in different 3d poses and views, and to save the corresponding 3d rig information as labels. We then train a deep neural network (DNN) to predict the rendered image from the 3d rig parameters. To allow dynamic lighting and scene composition, as well as produce high resolution images, we designed a progressive multi-branch generative network that outputs albedo, normals, mask and depth map. The different scene feature maps can be composed together with other scene elements (e.g.\ through a depth buffer test) to render the final image of the scene with the character in it.  At run-time, the application manipulates the rig parameters, such as the skeleton via traditional spline interpolation, and feeds the parameters to the network to produce the required scene feature maps in real-time.

One of the main challenges with image-based rendering are ghosting effects during interpolation and extrapolation to new views and movements. This is especially the case when undergoing large deformations, such as with articulated systems. We evaluate both pose and view generalization capabilities, and show results in our accompanying video. In a nutshell, our network architecture (section \ref{Network}) and data generation process (section \ref{DataGeneration}) achieves qualitatively good generalization for interpolating views, but remains challenged by extrapolating new movements, never seen in the training data. Before we discuss our approach in detail, let us now discuss relevant~work.

%% file: sec/RelatedWork.tex
\section{Related Work}
One of the main benefits our approach brings over traditional image-based rendering \cite{Debevec:1996,Levoy96,Buehler:2001, Schodel2010,Schodel2012, Carranza:2003, Germann2010,Xu:2011,Casas20144DVT,Volino2014} and more recent forms of learned rendering \cite{pix2pix2016, chan2018everybody, aberman2018deep, Liu2018Neural,aliaks2019textured}, is that we do not carry  intermediate geometry or pre-computed image dataset. 

\subsubsection*{Image-based rendering. }
Early work in image-based rendering focused on rendering new viewpoints of static objects and scenes undergoing rigid transformations, from a sparse set of images~\cite{Debevec:1996,Levoy96,Buehler:2001}. 

The idea of combining view-dependant texture maps with proxy geometry to interpolate sparse viewpoints came as early as 1996 \cite{Debevec:1996, Levoy96}. For static objects, a proxy shape with scene texture can efficiently interpolate camera views \cite{Debevec:1996}, while light fields parameterize the view space \cite{Levoy96} w.r.t.\ two parallel rectangles. These ideas were then extended to support unstructured geometries \cite{Buehler:2001}.  

In the case of characters and articulated systems, one needs to interpolate not only views but also pose. Early work would segment out actors from video, and re-enact them \cite{Schodel2010,Schodel2012}, whilst in the same view. To interpolate new views, \cite{Carranza:2003} introduced a character mesh with view-dependant texture blending, which was computed from multiple video footage.
Subsequent work seek to blend across poses \cite{Germann2010,Xu:2011,Casas20144DVT,Volino2014}. These methods employ a proxy mesh whose edges can be perceived and require a heavy memory footprint when keeping the video data. %In contrast, a deep neural network compresses succinctly the pose-dependant appearance space and shows promising generalization capabilities.

%In fact, we have seen in recent years a surge of work leveraging deep neural networks showing promising results in image-based re-rendering. 

In contrast, deep neural networks require much less memory as they succinctly compresses the pose-dependant appearance space, and have shown promising results recent years.

\FigArchitecture

\subsubsection*{Neural image-to-image translation using Adversarial Learning}
The core innovation behind these new methods is the ability to train a deep generative network in a semi-supervised manner (without paired correspondences) using adversarial learning \cite{Goodfellow2014}.  A set of images from one domain can be translated into a target domain \cite{pix2pix2016}, such as the image of a skeleton into the image of a person \cite{chan2018everybody, aberman2018deep}. 

This approach however is conditioned on flat 2d skeletons without proper 3d orientation information, leading to ambiguities between front and back for example. As a result, undesirable artefacts may appear as the limbs and body turn.  

The other issue with adversarial learning is the limitation to a rather low image resolution, such as $256 \times 256$. To reach higher resolutions, \cite{Karras2017} used an effective progressive training scheme, in which up-scaling convolutional blocks are progressively trained, one image resolution at a time.

Subsequent work seek to reach high quality, high resolution rendering, but by using paired images provided by the pre-rendering of proxy geometry. 

\subsubsection*{Image-to-image translation with a proxy mesh}
To circumvent both the limitation to low resolution images associated with adversarial learning, and to remove ambiguities due to flat 2d maps (e.g. 2d skeletons missing 3d surface orientation), methods have utilized paired images produced by first rendering coarse geometry \cite{Brualla2018,Liu2018Neural,thies2019deferred,aliaks2019textured,Meshry2019} and then translating to the corresponding real-world counter-part. 

The rendered geometry allows to condition the network with less ambiguous data: a UV map \cite{thies2019deferred}, a low resolution avatar with patterned clothing \cite{Liu2018Neural}, a rendered 3d skeleton \cite{aliaks2019textured}, or noisy captured albedo and normals \cite{Brualla2018,Meshry2019}. All these methods first render the 3d geometry and thus require creating and manipulating this intermediate object in their pipeline.  In contrast, we directly learn from rig parameters, to render high resolution, free of surface orientation artefacts images of articulated characters. 

\subsubsection*{Learned Scene Representations and Rendering}
Closest in spirit to our work is the approach of \cite{Eslami1204}, which introduced the idea of learning a rendering conditioned on a latent low dimension scene vector, and camera configuration.
Their work is focused on learning a compact scene representation from set of low resolution images, while we are focused on learning from artistically authored rig parameterazations and high resolution rendering of objects undergoing articulation.  

A similar and interesting line of work is that of \cite{Lombardi2019}, which learn a scene representation from a set of images via a differentiable volume ray caster. Their networks turn multiple images into a latent vector (modeled as a multivariate Gaussian distribution), which is decoded into a volume, which in turn is rendered via volumetric ray casting. The volume can be rotated and thereby generalize to new in-between views and motions. It would be interesting to see if such a volumetric representation could be learned from rig parameters, or if encoding the rig parameters into an identically and independently distributed latent vector (as with variational autoencoders) helps generalize better to new movements.

%% file: sec/Dataset.tex
\section{Dataset}
\label{DataGeneration}

Our approach requires a full image capture of the character in different poses and views. We take a 3d model of the character, together with a pose dataset defined in rig space $q$ (detailed bellow), and render each pose in many different camera views $c$. Note that we use the rendering software as is, and automate the posing and rendering via scripting. 

To avoid redundant pose information, we transform rig parameters $q$ into camera space: $\tilde{q} = c^{-1}q$. As rigs, we experimented with kinematic skeletons defined as positions and orientations, and produced all results using only the orientation of joints. We postulate other traditional rigs such as vertex-based elements, would be sufficient---insofar they are dense enough to model the orientation of each limb. Blend shapes would also work, as long as they are combined with pose information to account for articulation~and~viewing.

Our network takes as input rig parameters in view space $\tilde{q}$ and outputs feature maps $I = \{I_i\}$, such as the RGB albedo $I_a$ and occupancy $I_m$ mask of the character. To support dynamic lighting and scene composition with other scene elements, we need additional feature maps such as normals $I_n$  and depth $I_d$, as described in our network extension section~\ref{Extensions} and scene composition section~\ref{SceneComposition}.  Hence when rendering the character, the rendered maps $I$ are save together with the corresponding rig parameters $\tilde{q}$. 

In the event that a rig would be too high dimensional for practical real-time use, we propose to parameterize a kinematic skeleton (joint positions and orientaitons) w.r.t.\ to the high dimensional rig, and to save the reduced skeleton subspace $\tilde{q}$ instead of the full rig space. While this parameterization would require additional work for the setup, it remains less laborious than re-creating a full character mesh, rig, and re-animating each movement.

%% file: sec/Network.tex
\FigExtensions

\section{Rig2Image Network}
\label{Network}

Our network takes the rig parameters $\tilde{q}$ directly as input, and outputs an image $I$ of the character in the given rig's pose. The challenges with designing such a network are first convergence while learning to generate high resolution detailed images, and second to efficiently generate several scene maps $I$ such as $I=\{I_a, I_n, I_d, I_m\}$ albedo, normals, depth, mask, etc.
We first detail how to architect and train a model for generating a single high resolution image $I_a$, and detail in section~\ref{Extensions} how to efficiently extend the architecture to multiple~maps. 

To converge at learning high resolution images, we build upon previous explorations, by using the progressive training scheme of \cite{Karras2017}, and adapting the generative network to arbitrary rig parameters. 
We first map the rig parameters $\tilde{q}$ to a fixed sized feature vector  using fully connected layers. Then we increase the spatial resolution of the features using multiple convolutional blocks, each doubling the spatial resolution, as shown in Fig.~\ref{Architecture}. 

Training this entire model up to full resolution from scratch fails to converge, resulting in large artefacts such as missing limbs and large holes in the character. %We found empirically that the key to convergence is to first train the network reaching maximum $16 \times 16$ resolution features, and to then progressively train each subsequent block, one at the time. 
We found empirically that convergence is best reached by first training the network to a resolution of $16 \times 16$, and to then progressively train each subsequent block, one by one. 

As training loss $L\left(I, I_{gt}\right)$, we minimize the error between the predicted image $I$ and the ground truth image $I_{gt}$.  We used a weighted combination of $L2$ and $L1$, as follows:
\begin{equation}
\nonumber
L\left(I, I_{gt}\right)= (1-\alpha) \cdot L2\left(I, I_{gt}\right) + \alpha \cdot L1\left(I, I_{gt}\right),
\end{equation}
where $L1$ tends to help increase sharpness slightly, and $\alpha$ is set to $0.1$ in all our experiments. 

Training with these losses leads to images that can remain blurry. To increase details, we experimented with losses from related work such a saliency loss \cite{Brualla2018} and multi-scale losses such as the perceptual loss \cite{Johnson2016Perceptual}. We obtained marginal increase in detail with these experiments. The multi-scale perceptual loss, which comes from a pre-trained network on natural images, injects textures closer to natural images and is thereby better suited for organic, photo-realistic images. A perceptual loss generated from a network trained on synthetic images, artistically similar to the character, may offer a bigger benefit and is left for future experimentation.  

Now that we can successfully generate a high resolution image from rig parameters, we describe how to efficiently generate multiple scene maps requires for scene composition.

% and perhaps a network trained on synthetic characters would be better.

%As loss we experimented extensively with losses from related work such a saliency loss \cite{Brualla2018}, multi-scale perceptual loss \cite{Johnson2016Perceptual}.  We did not obtain a perceivable increase in quality, though perceptual loss seemed to produce more organic images, which would be well suited for photo-realistic renderings. 

\comment{
We observed the error on the rendered images compared to ground truth was correlated to edges. To attenuate this, we steer the attention \cite{zhang2018selfattention} of the network to edges by computing the \newjb{spatial } gradient of the rendered image $\nabla I_p$, and increase the previous loss $L_s$ by multiplying it with the error between predicted gradient and ground truth gradient, resulting in:

\begin{equation}
\nonumber
L_{ga} = L_s(I-p, I_{gt}) \cdot (1 + \lambda \|\nabla I_p - \nabla  I_{gt} \|^2 ),  
\end{equation}

where $\lambda$ is a hyper-parameter set to $0.5$ in our results. As a 

}

\FigResultsAllMaps
\subsection{Multi-map Network for Dynamic Scene Composition}
\label{MultiBranchNet}
To enable dynamic lighting, as well as dynamic composition with other scene objects, our network needs to be able to generate different feature maps such as normals $I_n$, depth $I_d$, mask $I_m$, etc., together with the usual albedo $I_a$, in an efficient manner. For example, training independent networks for each map, significantly increases training time, and quickly reaches hardware constraints such as GPU memory. Additionally, since these feature maps are similar by nature, it is likely that they share deeper representations that could be more efficiently encoded in a common network.

In fact, certain feature maps such as depth, normals and mask are correlated and can be computed from one another; e.g.\ normals can be computed from depth by computing gradients, and a mask is a simple binary test from either depth or normals. Hence, we designed an architecture that shares deep representations up to the $8 \times 8$ features, then branches into two separate branches: one for albedo and one dedicated to geometric information such as depth, normals and mask, as shown in Fig.~\ref{Extensions}. As before, we first train the network to a resolution of $16 \times 16$, and then progress in parallel on each branch, dropping the \emph{toRGB}, \emph{toDepth, toNormals} and \emph{toMask} each time we move to the next block; up to $1024 \times 1024$ in our experiments.  %Note that experimentation with a single branch for all features rendered results that did not fully disentangle the features, i.e.\ albedo looking like normals and vice versa.
Fig.~\ref{fig_results_all_maps} shows an example of the predicted feature maps.

Using the scene maps $I_a, I_n, I_d, I_m$, we now describe how to integrate our learned character rendering into existing scenes.

%% file: sec/SceneComposition.tex
\section{Scene Composition}
\label{SceneComposition}
At run-time, the application manipulates the rig parameters, for example by blending poses via a game controller, or via data from a motion capture system, as shown in Fig.\ref{Runtime}. The rig parameters $\tilde{q}$ are then fed to our network to generate the feature maps $I$. 

One way to use our method is to simply learn a fully lit character and re-render it in the same views as it was trained on. However, to be able to compose the character with other objects in the scene (e.g.\ moving in front and behind), produce dynamic lighting, we need to use our feature maps $I=\{I_a, I_n, I_d, I_m\}$ and compose normals with light sources and depth with the scene's depth buffer. 

However, to be able to move the character in the camera frame, we need to \emph{transform}, or warp the feature maps as to appear positioned at new locations in the scene, as well scale the depth values $I_d$ according to the displacements.

To create the appearance of being rendered at new positions, we treat the character as a 3d billboard moving with the root motion of the 3d character, while facing the camera. This will automatically take into account the image-space scale of the character. Since we trained our model with poses centered at the camera, we need to take into account off-center viewing effects on the pose. 

For off-center viewing, we compute the rotation $R$ between the ray passing through the \emph{center point} of the camera, and the ray passing through the \emph{center point} of the billboard displaced in the scene.  We apply this rotation to the pose $\tilde{q}$, resulting in a new pose $\hat{q} = R \, \tilde{q}$ that we feed to our network to generate the maps $\hat{I}$.

To scale the depth maps, we compute the scale $s$ between the image-space billboard dimensions at run-time and the image-space billboard dimensions at training time. We then scale the depth values of $\hat{I}_d$ with the inverse of $s$ as to increase depth as the billboard gets smaller, resulting in a new depth map $\bar{I}_d = s^{-1} \, \hat{I}_d$. Using the adjusted maps in image space, we perform additional pixel-wise operations to compute light intensity using the normals $\hat{I}_n$, and compose with other objects using depth maps $\bar{I}_d$.

\subsubsection*{Dynamic Lighting}
To demonstrate dynamic lighting effects, we use the normal maps and compose them with light sources in the scene by computing Phong shading \cite{Phong1975}. We sum for each camera, the dot product between the light source pointing direction and the normal vector. Examples with different light colors are shown in the accompanying video. 

\subsubsection*{Composition with other scene objects}
When other objects move in front of the character, we need to know which pixels should be replaced by the other object's color. We solve this problem with a traditional graphics technique, i.e.\ the depth buffer \cite{Catmull1974}. Most modern renders produce depths maps when rendering. Hence we can integrate with other systems via this depth buffer interface. The final pixel color is the map for which the $z$ value is the nearest from all the depth maps. Examples of this composition are shown in the accompanying video.

%% file: sec/Results.tex
\section{Evaluation}
We first describe the dataset we used for our results and comparisons. Using this dataset we were able to train a deep neural network to render articulated figures at high resolutions ($1024 \times 1024$) directly from rig parameters.
We then evaluate our model on a core challenge with learned rendering, that is the generalization to \emph{new views} and \emph{new motions}.  We describe bellow evaluations to both, and evaluations of each can seen in our accompanying video.

\FigFrontBack

\FigResultWithError

\subsubsection*{Dataset}
In each experiment, we used the same dataset. We trained a model to go from skeleton rig parameters to the rendered image at resolution ($1024 \times 1024$). We utilized a motion library comprised of $5$ minutes of small, but diverse motion clips sampled at $60$ Hz, amounting to $18$k frames. We sample views at a regular interval on a circular disc around the character. In order to compare generalizabilty to new views, we trained one model on $6$ views, sampled at $60$ degrees intervals, amounting to 108k images and $7$ Gigabytes (GB) of data, and one model on $60$ views, amounting to 1,080k images and $170$ GB of data. Note that our neural network does not change and remains at a constant size of $350$ megabytes.

\subsubsection*{Results on Dataset}

We first evaluate the performance over the training data. We can see that our network is capable of differentiating between front and back using the rig parameters, as shown in Fig.~\ref{frontback}. The accompanying video and Fig.~\ref{fig_results_errror_vis} show the error between the predicted image and the ground truth. We can observe the error being larger for limb extremities, high frequency details and edges around the character, as well as poses less represented in the dataset. For further poses see Fig.~\ref{Results} or the accompanying video.

\FigNewViews

\subsubsection*{New views}
To evaluate generalizing to new views, we synthesized $360$ view transformations that we apply to a single pose from our training data. The pose was previously seen during training, but the combination of pose and view not. We can see that the model trained on $6$ views exhibits significant non-linear morphing, but that the model trained on $60$ views yields perceptually smooth results. As a reference for comparison, we computed a standard linear interpolation of the $60$ views, and one can see in Fig.~\ref{NewViews}, as well as in our video that the interpolation remains noticeably ghosty. For a quantitative comparison see Fig.~\ref{fig_NewViewsDiagramm}, which clearly shows the benefit over the linear interpolation.

\subsubsection*{New motions}
To evaluate generalizability to new movements, we tested on a separate motion clip that was not part of the training set. We can see in the accompanying video how the network might loose a limb when a significantly different pose is presented. Note that our first trials included the root position in the rig, and the network learned correlations specific to this value. Hence generalization was very poor, which led us to learn only from orientations. Hence a rule of thumb is to avoid absolute world or frame-relative rig coordinates.  

One way we believe could improve generalizibilty to new motions is to train on more diverse poses, instead of continuous motion clips. While our movements are diverse, the individual poses close to one another as a result of being continuous movement clips, which is easier for the network to encode, then poses that would be further.

\FigNewViewsDiagramm

\subsubsection*{Architecture}
We experimented with alternative architectures. In section~\ref{MultiBranchNet}, we introduce a multi-branch network for albedo, normals, depth and mask that shares deep layers and features. We experimented with a single branch, but this resulted in fighting between feature maps; albedo would look like normals, or vice versa.  

Many works utilize skip connections to help kernels operating on the high resolution features recover localized structures. We tried adding skip connections from the rig parameters to further layers, but this did result in additional details. We believe that a differentiable splatting of the rig into an image, combined with skip connections could improve the rendering details.

\FigResults

%We learn a rendering centered in camera space. In practice we might want to render the character in a scene where the character is in a corner. This would require a slight rotation of the pose to account for the angled view, as a well as a distortion of the character consistent with the camera's lens. We believe these could be convincingly approximated with additional engineering.

%Our new approach to character rendering leads to new potential applications. For example we could capture movements from previous movie productions directly into rig parameters by learning the inverse of our rendering (image2rig). 

%% file: sec/Limitations.tex
\section{Limitations and Future Work}

In this work, we demonstrate for the first time a learned character rendering generated directly from the rig parameters at high resolution ($1024 \times 1024$). Unlike human faces, articulated figures undergo large deformations in image space and require powerful models to represent compactly. While we have made a step forward in this direction, our rendered images remain blurry and could be improved with additional details. 

When presenting our network with new poses, never seen before, we can sometimes see a limb fading, or morphing artefacts. One way to improve this could be to craft a more diverse dataset, and challenge the network a bit more during training--instead of showing continuous motion clips with each pose being similar to one another. An alternative could be to employ a differentiable \emph{splatting} function that would draw a character structure from the rig parameters, which could help the kernels improve the rendering locally on the body parts.

%% file: sec/Conclusion.tex
\section{Conclusion}

We were able, for the first time, to render an articulated character using a learned model operating directly on its animation rig parameters. Our approach is fully automated and can take a high resolution asset rendered in its environment and re-render it in real-time with a deep neural network running on a GPU. This now opens up new possibilities for live rig-space previs and exciting research opportunities in real-time rendering. For example, it would be interesting to learn representations that can be shared across similar characters. Or another line of work is stylization, and real-time blending of digital characters into real world videos.